# A Transport and Microwave Study of Superconducting and Magnetic RuSr$_2$EuCu$_2$O$_8$

M. Požek, A. Dulcic, D. Paar

*Department of Physics, Faculty of Science, University of Zagreb, P. O. Box 331, HR-10002 Zagreb, Croatia*

G. V. M. Williams

*2. Physikalisches Institut, Universität Stuttgart, D-70550 Stuttgart, Germany and The New Zealand Institute for Industrial Research, P.O. Box 31310, Lower Hutt, New Zealand.*

S . Krämer

*2. Physikalisches Institut, Universität Stuttgart, D-70550 Stuttgart, Germany.*

## ABSTRACT

We have performed susceptibility, thermopower, dc resistance and microwave measurements on RuSr$_2$EuCu$_2$O$_8$. This compound has recently been shown to display the coexistence of both superconducting and magnetic order. We find clear evidence of changes in the dc and microwave resistance near the magnetic ordering temperature (132 K). The intergranular effects were separated from the intragranular effects by performing microwave measurements on a sintered ceramic sample as well as on a powder sample dispersed in an epoxy resin. We show that the data can be interpreted in terms of the normal-state resistivity being dominated by the CuO$_2$ layers with exchange coupling to the Ru moments in the RuO$_2$ layers. Furthermore, most of the normal-state semiconductor-like upturn in the microwave resistance is found to arise from intergranular transport. The data in the superconducting state can be consistently interpreted in terms of intergranular weak-links and an intragranular spontaneous vortex phase due to the ferromagnetic component of the magnetization arising from the RuO$_2$ planes.

## Introduction

There have been a number of recent studies reporting the coexistence of superconductivity and magnetic order in $RuSr_2R_2Cu_2O_{10}$ and $RuSr_2RCu_2O_8$ where R=Gd or Eu [1-12]. These superconductors were originally synthesized by Bauernfeind *et al.* [13,14]. Most recent reports have focussed on $RuSr_2GdCu_2O_8$ which has a unit cell similar to that of the $YBa_2Cu_3O_7$ high temperature superconducting cuprate (HTSC) where there are two $CuO_2$ layers and one $RuO_2$ layer with the $CuO_2$ and $RuO_2$ layers being separated by insulating layers [5,8]. However, it is more complicated than $YBa_2Cu_3O_7$ in that there is a coherent rotation of the $RuO_6$ octahedra within domains extending up to 20 nm in diameter. Recent magnetization and muon spin rotation studies have shown that there exists a magnetic ordering transition at a temperature much greater than the superconducting transition temperature, $T_c$ [3,4]. Some studies have been interpreted in terms of ferromagnetic order arising from the Ru moment in the $RuO_2$ layers [3-10,12]. This generated considerable interest because ferromagnetic order and superconductivity are competing processes and can only coexist via some accommodation, for example by a spatial modulation of the respective order parameters or the formation of a spontaneous vortex phase. However, a recent powder neutron diffraction study on $RuSr_2GdCu_2O_8$ has shown that the low-field magnetic order is predominantly antiferromagnetic and it is confined to the Ru moments in the $RuO_2$ layers [11]. Furthermore, a magnetization study on $RuSr_2EuCu_2O_8$ has shown that, while the low-field magnetic order is predominantly antiferromagnetic, there is a small ferromagnetic component of unknown origin [15]. For higher applied magnetic fields there is a transition to a predominantly ferromagnetic state. The transition commences at ~4 kGauss and it is not complete even at 60 kGauss.

There are still a number of unanswered questions concerning the magnetic and electronic properties of $RuSr_2RCu_2O_8$. For example, the magnetization displays a decrease near 45 K in $RuSr_2GdCu_2O_8$ but the diamagnetic transition occurs at a lower temperature of 35 K [6]. This, along with the results from resistivity and heat capacity measurements has been interpreted in terms of a spontaneous vortex phase attributed to the low-field ferromagnetic component of the magnetization [12] (~0.15 $\mu_B$/Ru at 5 K [15]). The decrease in the magnetization, the decrease in the resistivity and the peak in the heat capacity near 45 K in $RuSr_2GdCu_2O_8$ have been attributed to a "thermodynamic superconducting transition" [6] where the diamagnetic transition is suppressed due to a spontaneous vortex phase [12]. In the case of $RuSr_2EuCu_2O_8$, the decrease in the susceptibility occurs near 32 K while the transition to the bulk diamagnetic phase commences below 12 K [15,16]. The width of the transition (~

20 K) is much broader than that observed in other HTSC and clearly requires further investigation.

Another question pertains to the extent of coupling between the Ru moments in the $RuO_2$ layers and the carriers in the $CuO_2$ layers. A magnetotransport study on unorientated $RuSr_2GdCu_2O_8$ ceramic samples [9] found evidence for magnetoresistence effects above and below the magnetic ordering temperature (~132 K). For temperatures above the magnetic ordering temperature the magnetoresistance decreases as the square of the applied magnetic field which has been attributed to the freezing out of spin-disorder scattering as the Ru moments become aligned with the field. However, for temperatures below the magnetic ordering temperature the magnetoresistance displayed an anomalous increase and then decrease with increasing applied magnetic field. The magnetoresistance above the magnetic ordering temperature was analyzed within the Zener, or s-d model to extract an exchange energy. The size of the deduced exchange energy is large and comparable to the energy of the superconducting gap. It was not possible from this study to determine if the $RuO_2$ layers contributed significantly to the electronic transport.

There are a number of unexpected structural and transport changes that occur for temperatures in the vicinity of the magnetic ordering temperature. For example, studies on $RuSr_2GdCu_2O_8$ ceramic samples report a decrease in the Hall coefficient [9]. Structural refinement studies on $RuSr_2GdCu_2O_8$ show that only the Cu-Cu bond length (i.e. the distance between the $CuO_2$ planes) and the Cu-O-Cu bond angle are affected by the magnetic order [8].

In this paper we report the results from a transport and microwave study of $RuSr_2EuCu_2O_8$ with the aim to address the questions above and improve the understanding of the ruthenate-cuprates.

## Experimental Details

The $RuSr_2EuCu_2O_8$ ceramic samples were prepared using the same synthesis conditions used to make $RuSr_2GdCu_2O_8$ [6]. The starting materials were $RuO_2$, $Eu_2O_3$, CuO and $SrCO_3$. The synthesis process involved (i) decomposing at 960 °C in air for 12 hours, (ii) sintering at 1010 °C in flowing $N_2$ for 10 hours, (iii) sintering at 1050 °C in flowing $O_2$ for 10 hours, (iv) sintering at 1055 °C in flowing $O_2$ for 10 hours, (v) sintering at 1060 °C in flowing $O_2$ for 7 days. The samples were ground after each processing step. The first step is required to suppress the $SrRuO_3$ impurity phase [14]. The last process is crucial for obtaining samples with high zero resistance superconducting transition temperatures. The samples were characterized using x-ray diffraction and there was no evidence of the ferromagnetic $SrRuO_3$ or the $Sr_2EuRuO_6$ impurity phases to within the ~2% detection limit.

The resistance was measured between 5 K and 300 K and variable temperature thermopower measurements were made between 10 K and 300 K. The minimum measurable resistance was $5 \times 10^{-6}$ $\Omega$. The ac susceptibility data was obtained on a sintered ceramic rod using a SQUID in zero applied magnetic field. The ac magnetic field was 0.05 Gauss and the frequency was 1 kHz. The dc susceptibility measurements were made using a SQUID and with an applied magnetic field of 100 Gauss.

The microwave measurements were made in an elliptical $_eTE_{111}$ cavity operating at 9.3 GHz. The sample was mounted on a sapphire sample holder and positioned in the cavity center where the microwave electric field is maximum. The temperature of the sample could be varied from liquid helium to room temperature while the body of the microwave cavity was kept at liquid helium temperature. This enabled us to achieve high Q-factors (about 20 000 for the unloaded cavity) and good thermal stability. The cryostat with the microwave cavity was placed in a superconducting magnet so that the sample could be exposed to a dc magnetic field of up to 80 kGauss. The changes in the microwave electrical conductivity of the sample induced by either temperature or magnetic field were detected by a corresponding change in the Q-factor of the cavity. The quantity $1/2Q$ represents the total losses of the cavity and the sample. The experimental uncertainty in the determination of $1/2Q$ was about 0.03 ppm. The details of the detection scheme are given elsewhere [17].

**Results and Analyses**

We present in figure 1 the zero-field ac susceptibility data from SQUID measurements on $RuSr_2EuCu_2O_8$. The three main features are (i) a peak in the susceptibility near 132 K, (ii) a sudden decrease in the susceptibility for temperatures less than ~32 K, and (iii) the onset of bulk diamagnetism below ~12 K. The decrease near ~32 K has been attributed to the onset of superconductivity and the lower temperature decrease at ~12 K has been attributed to the onset of the Meissner phase [16] which, by comparison with a study on $RuSr_2GdCu_2O_8$ [12], may be suppressed due to a spontaneous vortex phase. The peak near 132 K is due to the onset of predominately low-field antiferromagnetic order. However, there is a small ferromagnetic component with a remanent magnetization at 5 K of 0.05 $\mu_B$/Ru [15]. In the case of $RuSr_2GdCu_2O_8$ the small ferromagnetic component at 5 K is three times larger than that in $RuSr_2EuCu_2O_8$. The peak near 132 K seen in figure 1 has been shown to rapidly disappear with increasing magnetic field and is no longer present for magnetic fields greater than 2.5 kGauss [15].

The dc resistance and thermopower are plotted against temperature in figure 2. Both samples of $RuSr_2EuCu_2O_8$ (samples A and B) exhibit weakly pronounced maxima in the dc resistance near the magnetic transition temperature (~132 K). This feature is more clearly seen in the insert to figure 2a (lower curve) where we plot the derivative of the dc resistance. A similar peak is also weakly evident in well-annealed $RuSr_2GdCu_2O_8$ as can be seen by the dashed curve in figure 2a and the corresponding derivative (upper curve in the figure 2 insert). At lower temperatures the dc resistance of $RuSr_2EuCu_2O_8$ shows a semiconductor-like upturn followed by the onset of superconductivity at 32 K. The zero resistance state occurs for temperatures below ~12 K. There is a small increase in the zero resistance temperature for our best sample (sample B) and only a small reduction in the semiconductor-like upturn. This sample was prepared using the same process described in the Experimental Section but a different person prepared it. Previous studies on $RuSr_2GdCu_2O_8$ have shown that the semiconductor-like upturn and the zero resistance temperature are critically dependent on the sample processing [5,14]. A High resolution TEM study on $RuSr_2GdCu_2O_8$ has shown that prolonged thermal treatment at 1060 °C removes most of the multidomain structure, consisting predominantly of 90 degrees rotations, as well as significantly reducing the semiconductor-like upturn [5].

The thermopower from $RuSr_2EuCu_2O_8$ is plotted in figure 2b. We note that the room temperature thermopower for $RuSr_2EuCu_2O_8$ (~73 µV/K) is only slightly greater than that observed in $RuSr_2GdCu_2O_8$ (~60 µV/K). It has been concluded that the room temperature thermopower of $RuSr_2GdCu_2O_8$ is comparable to that of an underdoped high temperature superconducting cuprate [6]. A similar interpretation of the room temperature thermopower from $RuSr_2EuCu_2O_8$ would indicate that the hole concentration in the $CuO_2$ planes is only slightly less than that in $RuSr_2GdCu_2O_8$. A lower hole concentration in $RuSr_2EuCu_2O_8$ could also explain why $T_c$ (as determined from the peak in the resistance and the initial decrease in the susceptibility) is lower in $RuSr_2EuCu_2O_8$ (~32 K) when compared with $RuSr_2GdCu_2O_8$ (~45 K). We show in the insert to figure 2b that, similar to the resistance data plotted in figure 2a, the derivative of the thermopower changes markedly near the onset of magnetic ordering temperature. This change could be due to the magnetic transition or it could be fortuitous because the temperature dependence of the thermopower is remarkably similar to underdoped $YBa_2Cu_3O_{7-\delta}$ [18] and $La_{2-x}Sr_xCuO_4$. For example, $YBa_2Cu_3O_{6.47}$ has a room temperature thermopower that is comparable to $RuSr_2EuCu_2O_8$ and there is a broad maxima centered near 170 K.

It can be seen in figure 2b that the thermopower from $RuSr_2EuCu_2O_8$ is near zero for temperatures less than ~20 K. However, there is a significant decrease in the thermopower for temperatures less than ~53 K. This temperature is greater than the temperature where the decrease in the ac susceptibility and resistance are observed (~32 K). In the case of $RuSr_2GdCu_2O_8$, the thermopower begins to significantly decrease for temperatures less than ~66 K while the resistance decrease, the zero thermopower, the change in the dc susceptibility and the peak in the heat capacity are all observed near 45 K [6]. The origin of the initial decrease in the thermopower at a temperature which is ~21 K above the significant change in the susceptibility at ~32 K in $RuSr_2EuCu_2O_8$ and ~45 K in $RuSr_2GdCu_2O_8$ is not clear. However, this correlation would appear in indicate that it is intrinsic.

We present in figure 3 temperature dependences of $1/2Q$ for applied magnetic fields up to 80 kGauss. It has previously been shown that the contribution to the total $1/2Q$ due to the sample is a measure of the microwave resistance [19]. For thick samples, the microwave penetration depth is much less than the sample thickness and $1/2Q$ is the real part of the surface impedance of the material. It is proportional to the square root of the sample resistivity. When the sample thickness is smaller than the penetration depth, $1/2Q$ depends linearly on resistivity. In the present case, the ceramic samples are thick while individual grains range from thin to thick with respect to the microwave penetration depth.

The zero-field curve plotted in figure 3 shows the onset of superconductivity at 32 K, which is the same temperature where the dc resistance begins to decrease. However, at lower temperatures the microwave resistance continuously decreases (at least for temperatures at, and above, 5 K) in contrast to the dc case where the dc resistance is zero below 12 K. We show later that this may be due to a spontaneous vortex phase. An increasing applied magnetic field has a dramatic effect on the resistance below ~32 K. The mechanisms involve flux penetration in intergranular weak links and the formation of the vortex phase in the grains [20]. These features will be analyzed later on when the data from a powder sample is also presented.

It can be seen in figure 3 that the zero-field curve has a small peak at ~130 K similar to the dc resistance peak in figure 2a. This peak is more apparent in the insert to figure 3. The peaks in both the microwave resistance and the dc resistance occur near the magnetic ordering temperature.

It is apparent in figure 3 that the peak at ~130 K is rapidly suppressed by an applied magnetic field. It is interesting to look at the field dependence of the microwave resistance in more detail. The magnetic field-dependence of $1/2Q$ at 130 K is shown in figure 4. One can notice a rapid decrease in $1/2Q$ for low magnetic fields followed by a transition to a slow, almost linear, dependence at high fields. However, the decrease does not saturate even at 80 kGauss, the highest magnetic field in our measurement. The rapid decrease of the microwave resistance for low magnetic fields is observed only for temperatures near 130 K. For temperatures further away from 130 K one observes only a slow linear decrease of the microwave resistance with increasing applied magnetic field. This is clearly seen in the insert to figure 3 and the curves for 80 K and 200 K plotted in figure 4. One may conclude that the microwave resistance around 130 K contains two contributions with different field dependences. The narrow peak seen in the insert to figure 3 appears to be superimposed on a broad maximum extending to ±50 K away from the peak.

The data in figures 3 and 4 can be understood by noting that the microwave penetration depth depends on the effective conductivity of the medium. The highly conducting $CuO_2$ planes and poorly conducting $RuO_2$ planes act in parallel so that the latter make a negligible contribution to the effective intragranular conductivity. The only significant effect of the $RuO_2$ layer is to cause additional scattering via exchange coupling between the Ru moment and the conduction band carriers in the $CuO_2$ layers. Thus, a simple explanation for the decrease in the microwave resistance with increasing magnetic field is that the applied magnetic field is suppressing an additional scattering mechanism, which arises due to fluctuations of the Ru moment. It can be noticed that the narrow peak in the microwave resistance disappears with increasing magnetic field in a manner similar to the disappearance of the predominately low-field antiferromagnetic order in the $RuO_2$ planes [15]. The broad maximum could then be associated with the ferromagnetic behavior. Further evidence that the $RuO_2$ layers do not directly contribute to the conductivity can be seen in the dc resistance and thermopower. There are no dramatic changes in the dc resistance near the magnetic transition temperature as seen for example in $SrRuO_3$ [21,22].

The granularity of the sintered sample is important in interpreting the transport measurements. In both the dc and microwave measurements the current flows not only in the $CuO_2$ planes of individual grains but also across the intergranular medium. This is a connection in series so that the corresponding resistivities must be added. As a result, the intergranular medium makes a significant contribution to the total resistivity. It is important to disentangle the contributions from the intergranular and the intragranular conduction paths.

For this reason, we prepared a powder sample which was embedded in an epoxy to eliminate the intergranular conduction paths.

We show in figure 5a that the magnetic properties of the powder sample are similar to those of the ceramic sample. Here we plot the zero-field-cooled (lower curve) and field-cooled (upper curve) dc magnetization at 100 Gauss. By comparing figures 1 and 5a it can be seen that both the ceramic and powder samples have the same magnetic transition temperature and the same superconducting transition temperature

Obviously, dc resistance measurements are excluded on powder samples, but microwave measurements with induced currents in individual grains are feasible. We present in figure 5b the zero-field microwave resistance curve for the same sample as in figure 5a. As expected, the overall microwave resistance is smaller than that from a ceramic sample of a comparable size. More important is the observation that the resistance curve for the powder sample does not show the strong semiconductor-like upturn at temperatures below ~120 K which is present in sintered samples (c.f. figures 2 and 3). This is clear proof that the pronounced semiconductor-like upturn in the sintered samples is due to the intergranular conduction paths. Also, in figure 5b one can see that the small peak in the microwave resistance for the powder sample is present in the same form as in the sintered samples. Therefore, it appears to be an intrinsic property of the magnetic transition in the $RuSr_2EuCu_2O_8$ compound.

As mentioned above, the superconducting state in $RuSr_2EuCu_2O_8$ is complex and affects the ac susceptibility, thermopower, dc resistance and microwave resistance in different ways. For example, the ac susceptibility, dc resistance and the microwave resistance all decrease for temperatures below ~32 K. However, the ac susceptibility shows the onset of a diamagnetic transition below ~12 K, the dc resistance is zero below 12 K but the microwave resistance continually decreases for temperatures down to 5 K. In an attempt to understand the origin of this complex behavior, we measured the magnetic field dependence of $1/2Q$ at different temperatures below $T_c$. The resultant $1/2Q$ is plotted against applied magnetic field in Figure 6 for temperatures increasing from 5 K to 25 K. At temperatures just below $T_c$, the microwave resistance increases smoothly with the applied field. The curve at 25 K in figure 6 is representative of such a behavior. At lower temperatures one can see a progressive development of a narrow minimum centered at zero field. A small applied magnetic field considerably increases the microwave resistance. At 5 K we find that 90% of the rapid low-field rise is achieved at 1 kGauss. This initial increase is followed by a much slower rise at higher magnetic fields. Similar behavior is also seen in weak-linked ceramic samples of other high temperature superconducting cuprates [20,23,24]. Below the superconducting onset temperature (~32 K) the superconducting order parameter is formed first in the individual grains. Only at lower temperatures does the coupling between the grains become larger than $k_BT$ so that bulk superconductivity, and hence diamagnetic screening, occurs. A small magnetic field is sufficient to drive the intergranular weak links into the normal state and thus sharply increase the microwave absorption. For higher applied magnetic fields, there are an increasing number of vortices formed in the superconducting grains. The microwave current drives vortex oscillations, and this process contributes to the increasing microwave dissipation.

We have also measured the magnetic field dependence of the microwave resistance for the powder sample embedded in an epoxy resin. The insert to figure 6 shows the curve at 5 K. The powder sample exhibits only a remnant of the low-field sharp minimum. This is clear evidence that the sharp minimum in $1/2Q$ seen in the sintered samples is due to intergranular weak links. It appears that grinding the sample into powder and dispersing the grains in an epoxy removes most of the weak-links associated with grain to grain conduction paths.

We now return to the analysis of the dc resistance and microwave resistance curves below $T_c$. As mentioned earlier, the resistance transitions are very broad. This can not be due to impurity phases because any impurity phase is below the XRD detection limit (~2 %). Furthermore, we do not believe that the broad resistance transitions below $T_c$ could be due to weak links whose Josephson current gradually increases below $T_c$ until a superconducting path is fully established. A scenario based on weak links is not sufficient to explain the microwave resistance data on sintered *and* powder samples. In particular, $1/2Q$ decreases continually from ~32 K down to 5K for both the sintered and powder samples in the absence of an external applied magnetic field. Since the powder sample is practically free of intergranular weak links, one needs another mechanism to explain the broad resistance transition widths.

We show below that the spontaneous vortex phase model proposed for $RuSr_2GdCu_2O_8$ and $RuSr_2Gd_{2-x}Ce_xCu_2O_{10+\delta}$ [2,12,16] can account for the broad superconducting transitions observed in both the dc and microwave resistance data. In this model the spontaneous magnetization from the $RuO_2$ layers results in a local magnetic field that is greater than the lower critical field, $B_{c1}$, for temperatures greater than, $T_{svf}$ and less than $T_c$ [25]. The effect of a spontaneous vortex phase is to suppress the Meissner phase as mentioned earlier and as is apparent in figure 1. The zero dc resistance temperature will occur between $T_{svf}$ and $T_c$ and the temperature at which it occurs, $T_{irr}$, will depend on the value of the magnetic irreversibility field.

Unlike, the dc resistance data, the zero-field microwave resistance is finite below $T_c$ even in the absence of a spontaneous vortex phase. This can be understood by considering the frequency dependence of the complex conductivity $\tilde{\boldsymbol{s}}(T,\boldsymbol{w})=\boldsymbol{s}_1(T)-i\boldsymbol{s}_2(T,\boldsymbol{w})$ where the real part $\boldsymbol{s}_1(T)$ is due to quasiparticle excitations at finite temperatures and $\boldsymbol{s}_2(T,\boldsymbol{w})=\left(\boldsymbol{m}_0\boldsymbol{w}\boldsymbol{l}_L(T)^2\right)^{-1}$ is due to the superconducting fluid. It is apparent that $\boldsymbol{s}_2(T,\boldsymbol{w})$ decreases with increasing frequency and this will lead to finite microwave absorption below $T_c$. However, as the temperature is reduced below $T_c$, $\boldsymbol{s}_2$ will rapidly increase and $\boldsymbol{s}_1$ will decrease. The net effect will be a rapid decrease in the microwave absorption below $T_c$, which is observed in the HTSC [20,23,24,26]. For applied magnetic fields greater than $B_{c1}$ or in the presence of a spontaneous vortex phase there are additional losses due to vortices being driven by the induced microwave currents. This process occurs in both sintered and powder samples. The pinning of vortices for temperatures less than $T_{irr}$ will lead to zero dc resistance. However, at microwave frequencies the vortices can oscillate within the pinning wells and still give rise to a finite resistance below $T_{irr}$.

It is clear in figures 3 and 5 that the width of the superconducting transition as measured by the microwave resistance technique is significantly broader than that measured at zero frequency. Furthermore, the zero-field microwave superconducting transition width is broader than theoretically expected and the temperature dependence of the microwave resistance below $T_c$ does not follow that observed in other HTSC [20,23,24,26]. As mentioned above, we expect a rapid decrease in the zero-field microwave resistance below $T_c$. However, it is apparent in figure 5b that there is a linear decrease in the microwave resistance below $T_c$ and the low-temperature microwave resistance is significantly greater than zero. We believe that the simplest explanation is that there exists a spontaneous vortex phase.

The spontaneous vortex phase interpretation is further supported by the microwave resistance data at 80 kGauss and plotted in figure 5b (filled circles) for the powder sample. It is remarkable that the temperature dependence of the microwave resistance below $T_c$ is linear at zero applied field and with an applied field of 80 kGauss. This indicates that the mechanism responsible for the broad transition at 80 kGauss is likely to be the same for zero applied field. At 80 kGauss there are clearly vortices in the samples and hence it is reasonable to assume that the linear temperature dependence in zero applied field is due to a spontaneous vortex phase.

**Conclusion**

In conclusion, we have performed a susceptibility, thermopower, dc resistance and microwave study on $RuSr_2EuCu_2O_8$, which has been shown to exhibit the coexistence of superconductivity and magnetic order. We show that there are clear and well-defined changes in the transport and microwave data about the magnetic transition temperature (132 K). In particular, there is a narrow peak in both the dc and microwave resistance at the magnetic ordering temperature. It is superimposed on a broad maximum which extends approximately 50 K above and below the magnetic ordering temperature. The resistance in this region decreases with increasing magnetic field. A consistent interpretation of the data is that the conduction mechanism is dominated by the $CuO_2$ layers but the fluctuations of the magnetic order parameter in the $RuO_2$ layers affects the scattering rate of the carriers in the $CuO_2$ layers. It is also shown that most of the low-temperature semiconductor-like increase in the normal-state arises from intergranular transport. Below $T_c$ (32 K) we find evidence of numerous superconducting weak-links in the sintered sample which are all driven normal for magnetic fields greater than ~5000 Gauss at 5 K. These weak links are practically absent in powder samples. A consistent interpretation of both the dc and microwave resistance data can be made in terms of a spontaneous vortex phase.

**Acknowledgements**

We acknowledge funding support from the New Zealand Marsden Fund and the Alexander von Humboldt Foundation. We thank J. L. Tallon and C. Bernhard for providing and processing one of the samples (sample A).

**FIGURES**

**Figure 1:** Plot of the $RuSr_2EuCu_2O_8$ zero-field ac susceptibility against temperature for an ac field of 0.05 Gauss and a frequency of 1 kHz. The susceptibility has not been corrected for demagnetization effects.

**Figure 2:** (a) Plot of the dc resistance against temperature for two $RuSr_2EuCu_2O_8$ samples (A and B, solid curves) and a $RuSr_2GdCu_2O_8$ sample (dashed curve). The solid horizontal line is zero resistance. Insert: plot of the derivative of the resistance from $RuSr_2EuCu_2O_8$ (sample A and lower curve) and $RuSr_2GdCu_2O_8$ (upper curve). (b) Plot of the thermopower against temperature for $RuSr_2EuCu_2O_8$ (sample A). Insert: plot of the concomitant derivative of the thermopower from sample A.

**Figure 3:** Plot of 1/2Q against temperature for applied fields of 0 Gauss, 1 kGauss, 3 kGauss, 6 kGauss and 80 kGauss. Insert: plot of 1/2Q over an expanded temperature range. The arrows indicate increasing magnetic field.

**Figure 4:** Plot of 1/2Q against magnetic field for temperatures of 80 K, 130 K and 200 K. The solid line is a guide to the eye.

**Figure 5:** (a) Plot of the zero-field-cooled (lower curve) and field-cooled (upper curve) dc magnetization against temperature for the powder sample in an epoxy resin and for an applied magnetic field of 100 Gauss. (b) Plot of 1/2Q against temperature for the same sample as in (a) in the absence of an external applied magnetic field (solid curve) and for an applied magnetic field of 80 kGauss (filled circles). Note that the background level is estimated to be 34 ppm.

**Figure 6:** Plot of 1/2Q against magnetic field for temperatures of 5 K, 10 K, 15 K 20 K and 25 K for the ceramic sample. The arrow indicates increasing temperature. Insert: Plot of 1/2Q against magnetic field at 5 K for the powder sample diluted in epoxy resin. Note that the background level is estimated to be 34 ppm.

Figure 1
Phys. Rev. B

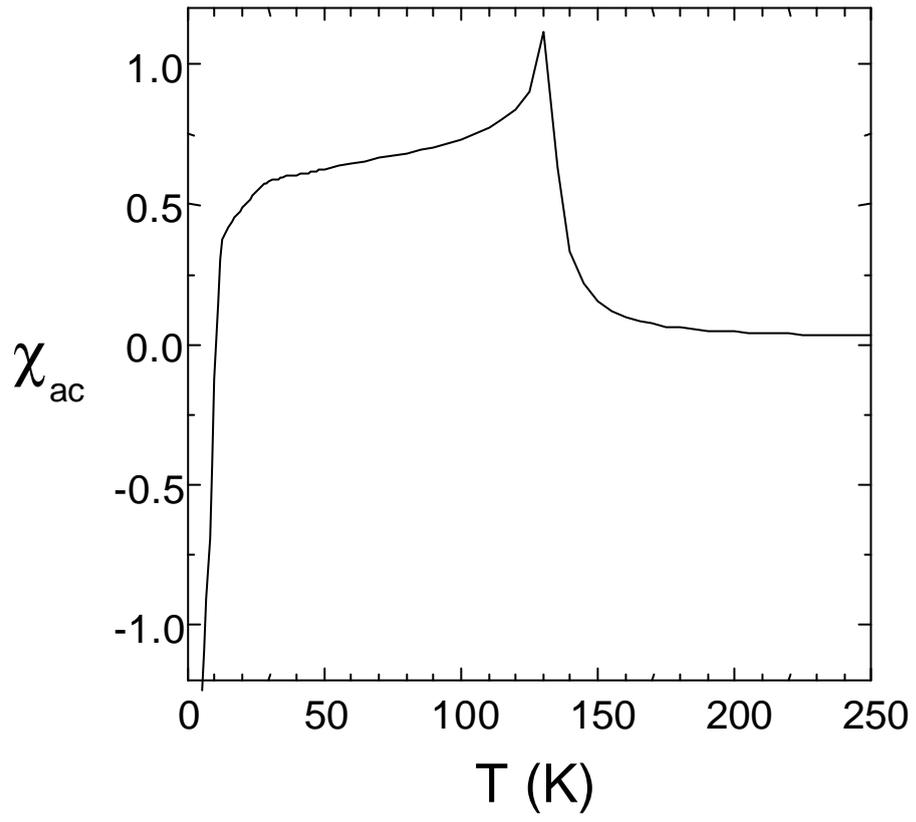

Figure 2
Phys. Rev. B

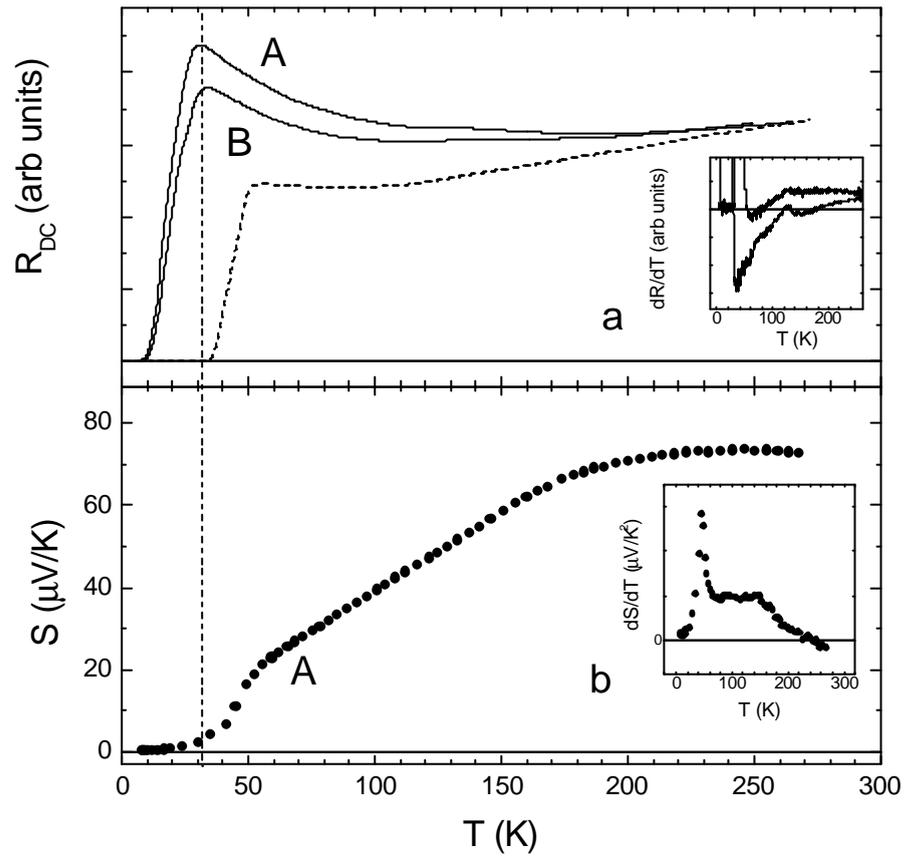

Figure 3
Phys. Rev. B

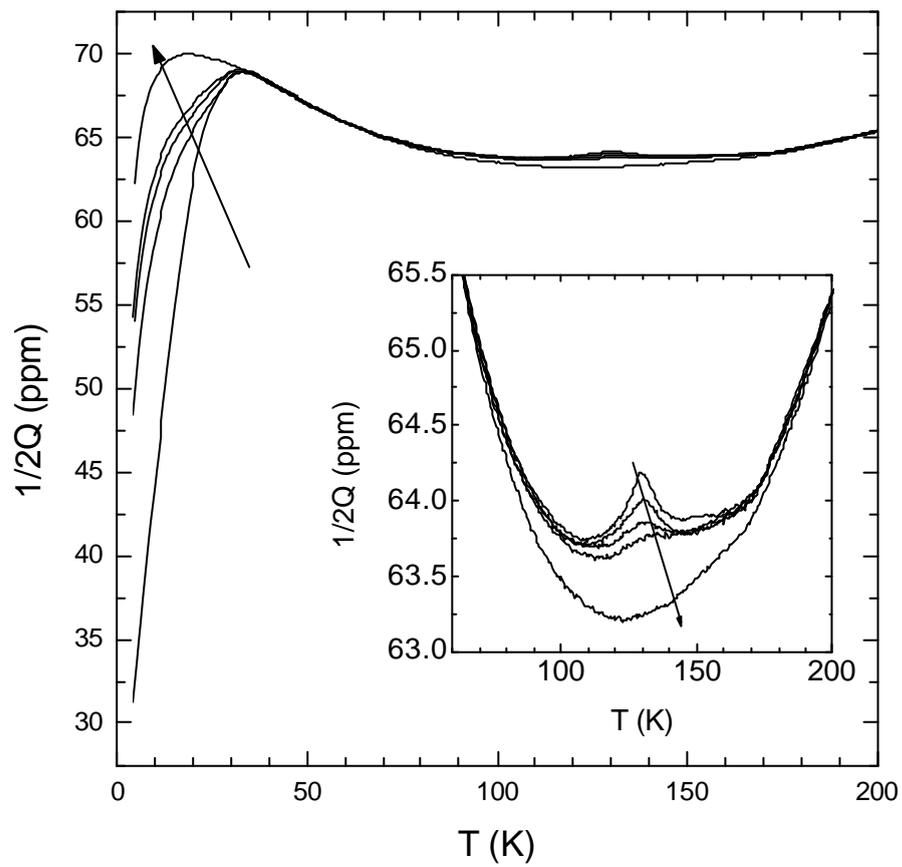

Figure 4
Phys. Rev. B

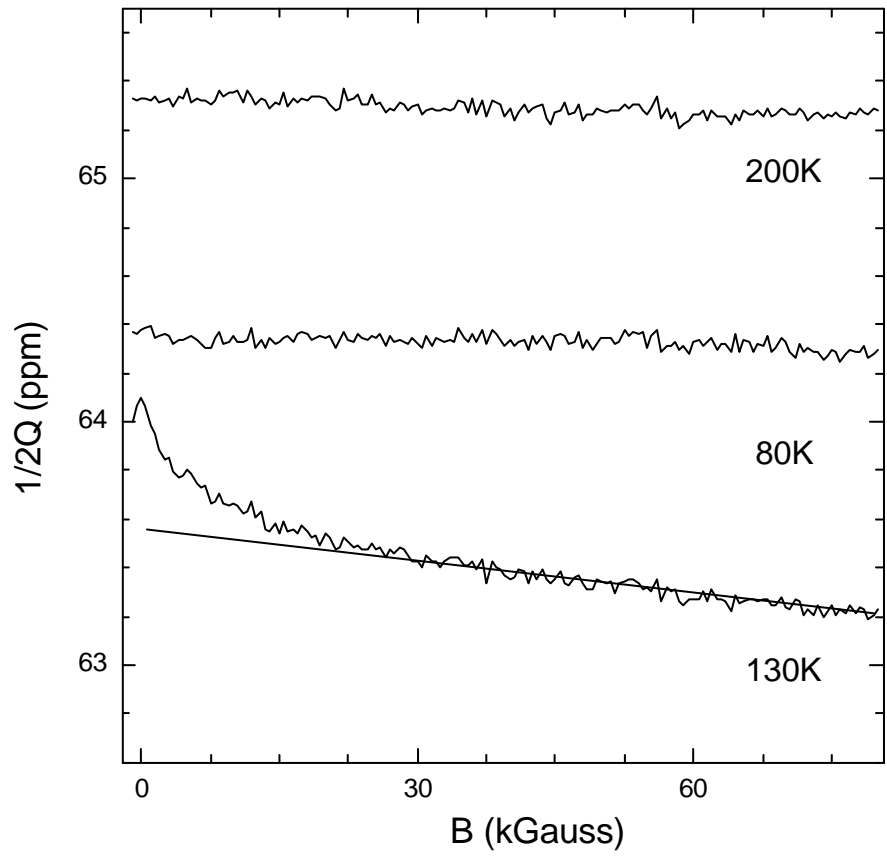

Figure 5
Phys. Rev. B

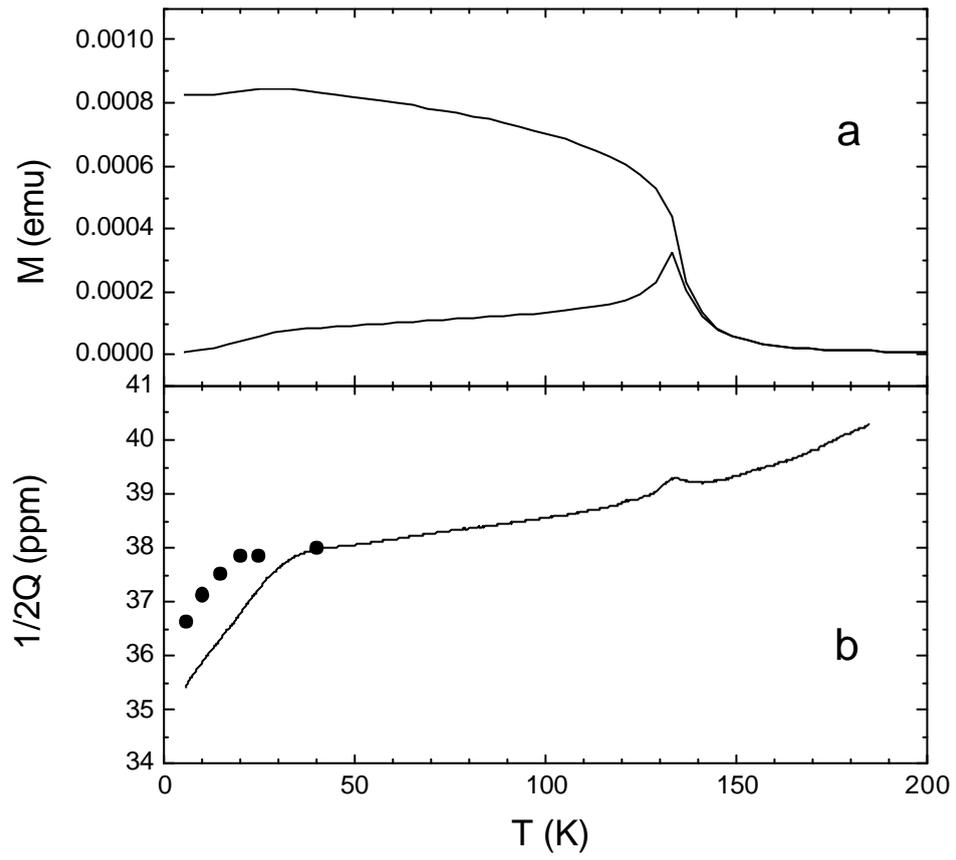

Figure 6
Phys. Rev. B

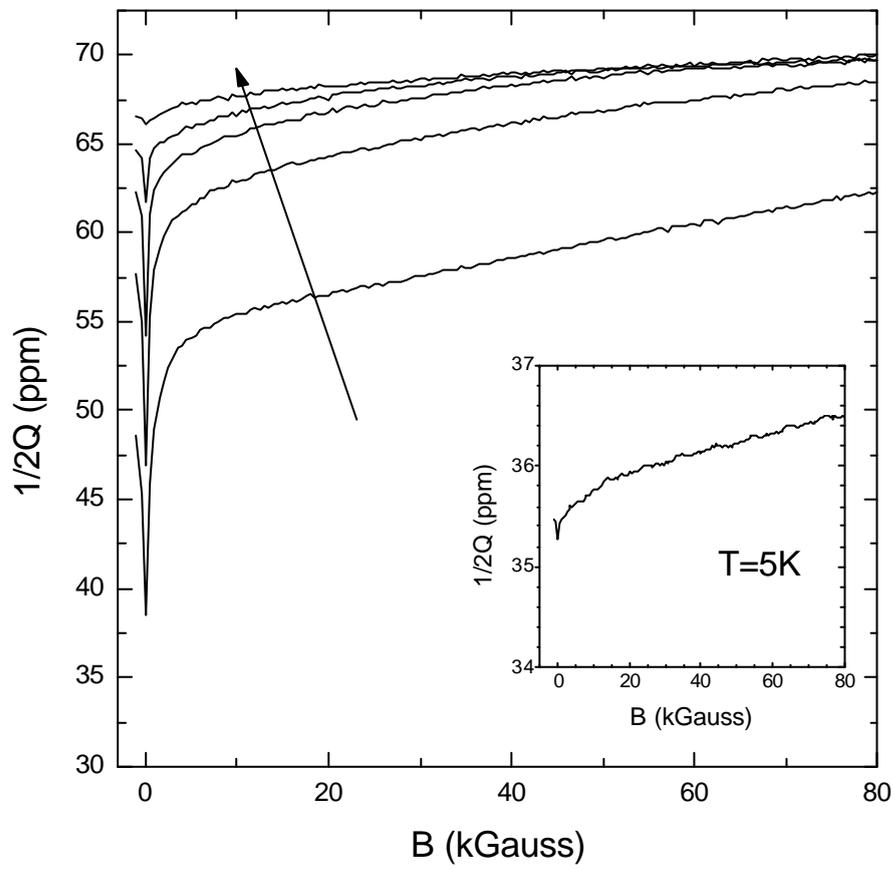